\documentclass[aps,pra,epsfigure,showpacs,twocolumn]{revtex4-1}
\usepackage{amsmath}
\usepackage{amstext}
\usepackage{latexsym}
\usepackage{graphicx}
\usepackage{amsfonts}
\usepackage{epsfig}
\usepackage{color}

\newcommand{\petr}[1]{\textcolor{black}{#1}} 

\begin{document}
\title{Elementary gates for quantum information with superposed coherent states}
\author{Petr Marek}

\affiliation{Department of Optics, Palack\'{y} University, 17. listopadu 1192/12, CZ-771 46 Olomouc, Czech Republic}

\author{Jarom\'{i}r Fiur\'{a}\v{s}ek}

\affiliation{Department of Optics, Palack\'{y} University, 17. listopadu 1192/12, CZ-771 46 Olomouc, Czech Republic}


\begin{abstract}
We propose a new way of implementing several elementary quantum gates for qubits in the coherent state basis. The operations are probabilistic and employ single photon subtractions as the driving force. Our schemes for single-qubit phase gate and two-qubit controlled phase gate are capable of achieving arbitrarily large phase shifts with currently available resources, which makes them suitable for the near-future tests of quantum information processing with superposed coherent states.
\end{abstract}

\maketitle
Quantum computation offers several advantages over its classical counterpart, namely an exponential speedup for some computational tasks. Currently, the most advanced approach to actually building the quantum computer relies on use of two level quantum systems - qubits. Their quantum optical implementation relied initially on states of single photons \cite{Qcomp}, but recently there were proposals to use superpositions of two `macroscopical' objects, two coherent states of light differing by phase \cite{kim02, ralph03,lund08}. Since then, there has been a considerable attention focused on obtaining such superposed coherent states \cite{catstates} or even arbitrary qubits in the coherent state basis \cite{catqubits}.

Any quantum computer needs to be constructed from basic building blocks, from quantum gates. In principle, two types of gates are required. Single-mode gates are needed to control quantum states locally, while two-mode gates serve to provide entanglement. The original proposal for quantum computing with coherent states \cite{ralph03} suggested that these gates could be implemented by coherent displacements and interference on unbalanced beam splitters followed by projection back onto the computational subspace. This approach  looks fine in theory, but with regards to currently available experimental resources, there is hardly any interesting effect that can be observed.

This statement requires some clarification. The scheme put forward in Ref. \cite{ralph03} relies on the phase shift that occurs when a coherent state gets displaced, $\hat{D}(\beta)|\alpha\rangle = e^{(\alpha\beta^{\ast}-\alpha^{\ast}\beta)/2}|\alpha+\beta\rangle$. If $|\beta|\ll |\alpha|$, the displaced state strongly resembles the original one, differing mainly in phase shift of the basis coherent state. The displacement could be driven classically, providing the single-mode phase-shift operation, or by another quantum state in order to implement a two-qubit gate. However, the need for the low value of the displacement results in a low value of the implemented phase shift, considering the currently achievable size of superposed coherent states, $|\alpha| \approx 1$. Consequently, a large number of operations (at least ten) would be required to achieve a $\pi$ phase shift.
 Furthermore, an indispensable part of the operation is quantum teleportation which projects the displaced state back onto the computational basis $|\alpha\rangle$, $|-\alpha\rangle$ and which should be implemented after each step. Without it, the actual nature of transformations is revealed to be that of a trivial displacement or a beam splitter. Unfortunately, the teleportation requires the entangled superposed coherent state as a resource, which, together with the need for photon number resolving detectors, renders it either unavailable or highly probabilistic.

All in all, the operations of \cite{ralph03} allow, in principle, deterministic interactions of arbitrary strength. In reality though, the single step produces only a very weak effect, and the need to teleport the states afterwards means that presently the full gate is probabilistic anyway and that there probably will not be more steps in the foreseeable future. \petr{Another option could be the teleportation based protocols of \cite{lund08}, which allow for arbitrarily strong effect and are, in principle, scalable. Here, only a single teleportation operation is required and the nonlinear effect is hidden in the specially prepared entangled resource state. Unfortunately, to prepare this state one needs access to perfect photon number resolving detectors, which are not available at this point.}Therefore, if we wish to test the principles of quantum information processing with superposed coherent states \petr{any time soon}, we need to devise alternative, more feasible, approaches.

In the following, we are going to present an alternative way of performing several of the elementary gates - the single-mode phase gate, the two-mode controlled phase gate, and the single-mode Hadamard gate. The gates are probabilistic, relying on projective measurements (photon subtractions, in particular) to deliver the non-linear effect.

\begin{figure}
\centerline{\psfig{figure = 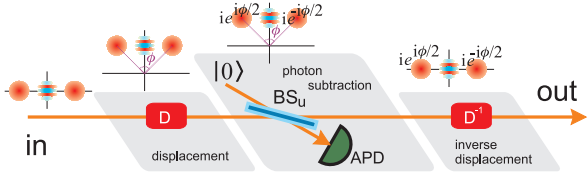,width=0.9\linewidth}}
\caption{(Color online) Schematic representation of the single-mode phase gate. BS stands for a mostly transmitting strongly unbalanced beam splitter, APD stands for avalanche photodiode, and D represents the displacement operation.}\label{fig_setup1}
\end{figure}

In order to clearly convey the basic ideas, let us work in the idealized scenario of perfect superposition of coherent states and perfect photon subtraction. We start with the single-mode phase gate which is necessary for single qubit manipulations. The procedure is schematically shown in Fig.~\ref{fig_setup1}.
An arbitrary qubit in the coherent state basis,
\begin{equation}\label{psiin}
    |\psi_{\mathrm{in}}\rangle=x |\alpha\rangle + y |-\alpha\rangle,
\end{equation}
is first coherently displaced by $\gamma$,
$ |\psi_{\mathrm{in}}\rangle\rightarrow \hat{D}(\gamma)|\psi_{\mathrm{in}}\rangle$. This operation can be easily performed by mixing the signal beam with an auxiliary strong coherent field on a  highly unbalanced beam splitter \cite{paris}.
Subsequently, a single photon is subtracted from the state, which is mathematically described by
the action of annihilation operator $\hat{a}$.
Finally, the state undergoes an inverse displacement by $-\gamma$, and we have
\begin{eqnarray}\label{}
 |\psi_\mathrm{out}\rangle&=&\hat{D}(-\gamma) \hat{a}\hat{D}(\gamma)|\psi_{\mathrm{in}}\rangle \nonumber \\
   &= &x (\alpha +\gamma) |\alpha\rangle + y (-\alpha  + \gamma) |-\alpha\rangle.
\end{eqnarray}
This operation becomes equivalent to a phase gate provided that the complex displacement $\gamma$ satisfies
\begin{equation}
\frac{\gamma-\alpha }{\gamma+\alpha }= e^{i\phi},
\end{equation}
which yields $\gamma=i\alpha/\tan(\phi/2)$. The output state after phase gate then reads
\begin{equation}\label{}
  |\psi_{\mathrm{out}}\rangle = i(x e^{-i\phi/2} |\alpha\rangle + y e^{i\phi/2} |-\alpha\rangle),
\end{equation}
and it can be seen that, up to a global phase factor, any nonzero phase shift $\phi$ may be performed in this way. 

\begin{figure}
\centerline{\psfig{figure = 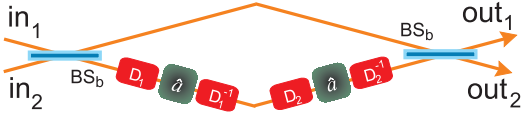,width=0.8\linewidth}}
\caption{(Color online) Schematic representation of the two-mode controlled phase gate. BS$_{\mathrm{b}}$ stands for a balanced beam splitter and D$_{1,2}$ represent displacements by $\gamma_{1,2}/\sqrt{2}$. Numbers 1 and 2 distinguish the two participating modes, while labels `in' and `out' describe the input and output state of the gate.}\label{fig_setup2b}
\end{figure}
Another important gate for quantum information processing is the two-qubit controlled phase gate, which is, up to local operations, equivalent to C-NOT gate, and which is used to establish entanglement in cluster states.
It can be implemented in a manner similar to the single-qubit phase gate, also employing displacements and photon subtractions as the driving force. However, to achieve interaction between the two modes $1$ and $2$ while preserving the computational basis, the operations take place in one arm of a Mach-Zehnder interferometer, see Fig.~\ref{fig_setup2b}.
For the input two-qubit state written in coherent state basis
\begin{equation}\label{PsiCPin}
|\Psi_\mathrm{in}\rangle=   c_{11} |\alpha,\alpha\rangle + c_{10}|\alpha,-\alpha\rangle +c_{01}|-\alpha,\alpha\rangle + c_{00}|-\alpha,-\alpha\rangle,
\end{equation}
the controlled phase gate is symmetric and preserves the structure of the state, only providing the term $|-\alpha\rangle|-\alpha\rangle$ with a phase factor $e^{i\phi}$, where $\phi$ is the phase shift introduced by the gate. A normalized output state of the gate corresponding to input state (\ref{PsiCPin}) thus reads
\begin{equation}\label{}
|\Psi_\mathrm{out}\rangle=   c_{11} |\alpha,\alpha\rangle + c_{10}|\alpha,-\alpha\rangle +c_{01}|-\alpha,\alpha\rangle + e^{i\phi}c_{00}|-\alpha,-\alpha\rangle,
\end{equation}
which is a new state with coefficients $c'_{mn}$ related to $c_{mn}$ as
\begin{equation}\label{cphase}
\frac{c_{11}'}{c_{11}}=\frac{c_{01}'}{c_{01}}=\frac{c_{10}'}{c_{10}}=\frac{c_{00}'}{c_{00}} e^{-i\phi}.
\end{equation}
 The implementation of the gate requires a Mach-Zehnder interferometer with two single photon subtractions accompanied by suitable displacements placed in one of the arms, which can be formally expressed as
\begin{eqnarray}\label{}
   |\Psi_{\mathrm{out}}\rangle &=& \hat{U}_{BSb}^{\dag}\hat{D}_2^{\dag}\hat{a}\hat{D}_2\hat{D}_1^{\dag}\hat{a} \hat{D}_1\hat{U}_{BSb}|\Psi_{\mathrm{in}}\rangle \nonumber \\
    &=& (\hat{a}+\hat{b}+\gamma_2)(\hat{a}+\hat{b}+\gamma_1)|\Psi_{\mathrm{in}}\rangle.
\end{eqnarray}
Here, $\hat{a}$ and $\hat{b}$ represent annihilation operators of modes $1$ and $2$, respectively, $\hat{D}_{1,2}$ stand for the displacement operators acting as $\hat{D}_{1,2}^{\dag}\hat{a}\hat{D}_{1,2} = \hat{a} + \gamma_{1,2}/\sqrt{2}$, and $\hat{U}_{BSb}$ is the unitary evolution operator of a balanced beam splitter, $\hat{U}_{BSb}^{\dag}\hat{a}\hat{U}_{BSb} = (\hat{a} + \hat{b})/\sqrt{2}$.

After the transformation, the composition of the state remains the same, only the coefficients are transformed to
\begin{eqnarray}
  c_{11}' &=& c_{11} (4 \alpha^2 +2\alpha(\gamma_1+\gamma_2) + \gamma_1\gamma_2), \nonumber \\
  c_{10}' &=& c_{10} \gamma_1\gamma_2, \nonumber \\
  c_{01}' &=& c_{01} \gamma_1\gamma_2, \nonumber \\
  c_{00}' &=& c_{00} (4\alpha^2 - 2\alpha(\gamma_1 +\gamma_2) + \gamma_1\gamma_2).
\end{eqnarray}

In order to achieve the controlled phase gate transformation given by (\ref{cphase}) one needs to attune the displacements $\gamma_1$ and $\gamma_2$ in such a way that
\begin{eqnarray}\label{}
    \gamma_1 + \gamma_2 = -2\alpha, \nonumber \\
    \gamma_1 \gamma_2 = \frac{8\alpha^2}{e^{i \phi} -1}.
\end{eqnarray}
An explicit calculation provides closed analytical formulas for the required displacements,
\begin{equation}
\gamma_{1,2}= -\alpha \left[1 \pm \sqrt{\frac{e^{i\phi}-9}{e^{i\phi}-1}}\right].
\end{equation}
Again, the phase shift $\phi$ can attain an arbitrary nonzero value.

\begin{figure}
\centerline{\psfig{figure = 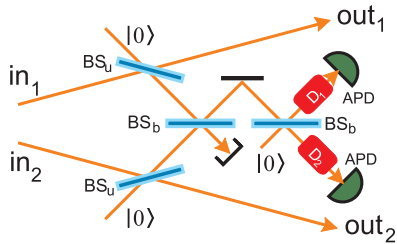,width=0.6\linewidth}}
\caption{(Color online) Alternative architecture of the two-mode controlled phase gate. BS$_{b}$ stands for a balanced beam splitter, while BS$_{u}$ represents a strongly unbalanced weakly reflective one. APD represents avalanche photodiode and D$_1$ and D$_2$ are the displacement operations. Numbers 1 and 2 distinguish the two participating modes, while labels `in' and `out' describe the input and output state of the gate.}\label{fig_setup2}
\end{figure}

It is important to stress, and it holds for both the phase gates, that although we have used direct displacements of the participating modes, it is actually more feasible to apply all the required displacement operations only on the ancillary modes used for the photon subtraction, just before the APD measurement. To explain the procedure we consider an arbitrary two-mode coherent state $|\alpha',\beta'\rangle$ and subject it to the evolution sketched in Fig.~\ref{fig_setup2}. First, the two modes are separately split on strongly unbalanced beam splitters with transmission coefficients $t \approx 1$ and reflection coefficients $r \ll 1$, which leads to a joint state
$|\alpha',\beta'\rangle |r\alpha',r\beta'\rangle$. The two ancillary modes are now mixed on a balanced beam splitter and one of the modes is traced out. Since $r$ is very small, this does not significantly reduce the purity and we can keep working with the state vector. The remaining mode is then split on another balanced beam splitter and two displacement operations are performed, arriving at the pre-measurement state
\begin{equation}\label{}
    |\alpha',\beta'\rangle |\frac{r}{2}(\alpha'+\beta')+\gamma_1',\frac{r}{2}(\alpha'+\beta')+\gamma_2'\rangle.
\end{equation}
In the limit of small $r$, the APD detectors can be represented by projection onto the single-photon Fock state $\langle 1|$ and if the displacements are chosen so $\gamma_{1,2}' = \gamma_{1,2} r/2$ the final state looks as
\begin{equation}\label{}
    (\alpha'+\beta' + \gamma_1)(\alpha'+\beta'+\gamma_2)|\alpha',\beta'\rangle,
\end{equation}
which is exactly what we want.

Note that a similar approach was already used for generation of an arbitrary coherent state qubit \cite{catqubits}, which also demonstrated, albeit in a limited way, the core principle behind the single-mode phase gate.

Finally, to complete the set of gates necessary for implementation of an arbitrary single qubit operation, we present a feasible implementation of Hadamard gate. Unlike the two previous gates, the Hadamard gate requires more than single photon subtractions. This is quite understandable, because the gate is supposed to transform a coherent state $|\alpha\rangle$ into a superposed state $|\alpha\rangle + |-\alpha\rangle$, which is a strongly non-linear process. Therefore an additional superposed coherent state, let's say $|\alpha\rangle + |-\alpha\rangle$, is required.

The core principle is simple and it employs the previously described controlled phase gate. This gate, with $\phi = \pi$, transforms the initial and the ancillary state to
\begin{equation}\label{}
    x|\alpha\rangle(|\alpha\rangle + |-\alpha\rangle) + y |-\alpha\rangle (|\alpha\rangle - |-\alpha\rangle).
\end{equation}
The gate is finalized by using a projective measurement $\langle \pi|$ such that $\langle\pi|\alpha\rangle = \langle \pi|-\alpha\rangle$. An example of such a measurement is the homodyne detection of the $\hat{p}$ quadrature, post-selecting the state only if a specific value is detected, or a photon number resolving detector projecting on an arbitrary even number Fock state.

\begin{figure}
\centerline{\psfig{figure = 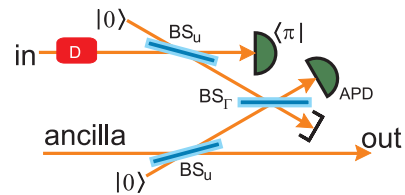,width=0.6\linewidth}}
\caption{(Color online) Schematic representation of the approximate single-mode Hadamard gate. BS$_{u}$ stands for a highly unbalanced weakly reflecting beam splitter, while BS$_{\Gamma}$ is a beam splitter with transmission coefficient $t_{\Gamma}$ used to set the parameter $\Gamma$. APD stands for a avalanche photodiode and $\langle \pi |$ represents the suitable projective measurement (see text).}\label{fig_setup3}
\end{figure}

This kind of Hadamard gate requires three projective operations. Two photon subtractions for implementation of the controlled gate and one additional measurement to confine the state into a single mode. There is another possibility, illustrated in Fig.~\ref{fig_setup3}, which reduces the number of operations to two. This improvement is compensated by imperfection of the operation, as it works only approximatively, even though the quality may be made arbitrarily large.

Here too we need another superposed coherent state $|\alpha\rangle + |-\alpha\rangle$. If we consider a displacement by some amplitude $\beta$, a single photon subtraction, and the inverse displacement, the state would be transformed to
\begin{equation}\label{}
    (\alpha+\beta) |\alpha\rangle + (-\alpha + \beta)|-\alpha\rangle.
\end{equation}
We can now see that for $\beta = 0$ we have obtained an odd cat state, while for $\beta \gg \alpha$ the cat state remained even. If we could correlate the displacement with the basis states of the initial state $x|\alpha\rangle + y|-\alpha\rangle$, we would have obtained the required transformation. So how to do it?

Let us start with the initial state (\ref{psiin}) and displace it by $\alpha$. The complete state of the initial mode and the resource mode then looks as
\begin{equation}\label{}
    (x|\beta\rangle+ y|0\rangle)\otimes (|\alpha\rangle+|-\alpha\rangle),
\end{equation}
where $\beta = 2\alpha$  but its value could be different if the initial state had a different amplitude than the ancillary resource. The next step is to apply a joint single photon subtraction, similarly as for the controlled phase gate, represented by operator $\Gamma \hat{a} + \hat{b}$ (where $\hat{a}$ and $\hat{b}$ are annihilation operators acting on the ancillary and the input mode, respectively) and a projection of the initial mode onto a certain pure state $\langle \pi|$ that will be specified below. The resulting single-mode output state then reads
\begin{equation}
x\langle \pi|\beta\rangle [(\beta+\Gamma\alpha)|\alpha\rangle+(\beta-\Gamma\alpha)|-\alpha\rangle] + y\langle\pi|0\rangle \Gamma \alpha (|\alpha\rangle-|-\alpha\rangle).
\end{equation}
If $|\Gamma\alpha|\ll |\beta|$ holds, we can make approximation $\beta\pm\Gamma\alpha \approx \beta$ and the output state simplifies to
\begin{equation}
x \langle \pi |\beta\rangle  \beta (|\alpha\rangle+|-\alpha\rangle) + y\langle\pi|0\rangle \Gamma \alpha (|\alpha\rangle-|-\alpha\rangle).
\end{equation}
The desired Hadamard operation is then performed if
\begin{equation}\label{}
    \langle \pi|\beta\rangle \beta  = \langle\pi| 0\rangle \Gamma \alpha.
\end{equation}
To achieve this, the projective measurement $|\pi\rangle$ needs to be properly chosen. For example, using homodyne detection to project on a $\hat{x}$ eigenstate $\langle \hat{x}=q|$ is appropriate, provided that $ \exp[- (q-\sqrt{2}\beta)^2/2] = \exp(-q^2)\alpha \Gamma/\beta$.  This can always be done.
The value of $\Gamma$ itself can be set by manipulating the beam splitter of the joint photon subtraction as $\Gamma = t_{\Gamma}/\sqrt{1-t_{\Gamma}^2}$. In this way, even if there is a large difference in amplitudes of the two participating states, the Hadamard gate can be implemented with arbitrary precision. Note that the standard way of generating an odd superposed coherent state by a photon subtraction is actually very close to implementation of the proposed Hadamard gate for a known coherent state input.

The experimental implementation of the proposed gates should be straightforward. The most difficult part of the gates is the photon subtraction, which can be implemented by a strongly unbalanced beam splitter and an on-off photo-detector - the avalanche photodiode. In this form the photon subtraction is becoming a staple of continuous variables quantum optical experiments and it is widely used to generate superposed coherent states \cite{catstates,catqubits}, or to manipulate and concentrate entanglement \cite{opatrny00,entdist}.

To summarize, we have proposed a feasible implementation of several elementary gates for superposed coherent state qubits. The main benefit of the proposed approach, which is based on using single photon subtractions, is that it allows achieving strong nonlinearities even with currently available small cat-like states exhibiting $|\alpha| \approx 1$ \petr{and without the need for photon number resolving detectors}. \petr{In fact, the proposed gates can be also used for generation of the resource states for the scalable teleportation based gates of \cite{lund08}.} The experimental feasibility, together with the ability to produce strong nonlinearities, makes these gates suitable for immediate tests of quantum information processing with coherent state qubits.

\medskip
\noindent {\bf Acknowledgments}
This research has been supported by projects  MSM 6198959213, LC06007 and Czech-Japan Project ME10156 (MIQIP) of the Czech Ministry
of Education. We also acknowledge grant 202/08/0224 and P205/10/P319 of GA CR and and EU grant FP7 212008 - COMPAS.


\begin{thebibliography}{9}
\bibitem{Qcomp}
E. Knill, L. Laflamme, and G.J. Milburn, Nature London \textbf{409}, 46 (2001);
R. Raussendorf and H. J. Briegel, Phys. Rev. Lett. \textbf{86}, 5188 (2001);
H. J. Briegel, D. E. Browne, W. D\"{u}r, R. Raussendorf, and M. V. den Nest, Nature Physics \textbf{5}, 19 (2009).

\bibitem{kim02}
H. Jeong and M. S. Kim, Phys. Rev. A \textbf{65}, 042305 (2002).

\bibitem{ralph03}
T.C. Ralph, A. Gilchrist, G.J. Milburn, W.J. Munro, and S. Glancy, Phys. Rev. A \textbf{68}, 042319 (2003).

\bibitem{lund08}
\petr{A.P. Lund, T. C. Ralph, and H. L. Haselgrove, Phys. Rev. Lett. 100, 030503 (2008).}

\bibitem{catstates}
A. Ourjoumtsev, H. Jeong, R. Tualle-Brouri, and P. Grangier, Nature \textbf{448}, 784 (2007);
J. S. Neergaard-Nielsen, B. M. Nielsen, C. Hettich, K. M{\o}lmer, and E. S. Polzik, Phys. Rev. Lett. \textbf{97}, 083604 (2006);
H. Takahashi, K. Wakui, S. Suzuki, M. Takeoka, K. Hayasaka, A. Furusawa, M. Sasaki, Phys. Rev. Lett. \textbf{101}, 233605 (2008);
T. Gerrits, S. Glancy, T.S. Clement, B. Calkins, A.E. Lita, A.J. Miller, A.L. Migdall, S.W. Nam, R.P. Mirin, and E. Knill, arXiv:1004.2727.

\bibitem{catqubits}
J. S. Neergaard-Nielsen, M. Takeuchi, K. Wakui, H. Takahashi, K. Hayasaka, M. Takeoka, M. Sasaki, arXiv:1002.3211.

\bibitem{paris}
M.G.A. Paris, Phys. Lett. A \textbf{217}, 78 (1996).


\bibitem{opatrny00}
T. Opatrn\'{y}, G. Kurizki, and D.-G. Welsch, Phys. Rev. A \textbf{61}, 032302 (2000).


\bibitem{entdist}
H. Takahashi, J. S. Neergaard-Nielsen, M. Takeuchi, M. Takeoka, K. Hayasaka, A. Furusawa, M. Sasaki, Nature Photonics \textbf{4}, 178 (2010).


\end{thebibliography}
\end{document}